\documentclass{osa-article}

\journal{oe}

\articletype{Research Article}
\usepackage{mathrsfs} 
\usepackage{amsmath,amssymb}
\usepackage{booktabs}
\usepackage{lipsum}
\usepackage{mathtools}
\usepackage{stackrel}
\begin{document}

\title{Low Photon Budget Phase Retrieval with Perceptual Loss Trained Deep Neural Networks}

\author{Mo Deng,\authormark{1,*} Alexandre Goy,\authormark{2,*} Shuai Li,\authormark{2}, Kwabena K. Arthur,\authormark{2} and George Barbastathis \authormark{2,3}}

\address{\authormark{1}Department of Electrical Engineering and Computer Science, Massachusetts Institute of Technology, Cambridge, MA02139, USA\\
\authormark{2}Department of Mechanical Engineering, Massachusetts Institute of Technology, Cambridge, MA02139, USA\\
\authormark{3}Singapore-MIT Alliance for Research and Technology (SMART) Centre, Singapore 117543, Singapore}

\email{\authormark{*}modeng,agoy@mit.edu} 

\begin{abstract}
Deep neural networks (DNNs) are efficient solvers for ill-posed problems and have been shown to outperform classical optimization techniques in several computational imaging problems. DNNs are trained by solving an optimization problem implies the choice of an appropriate loss function, \textit{i.e.} the function to optimize. In a recent paper [A. Goy \textit{et al.}, Phys. Rev. Lett. 121(24), 243902 (2018)], we showed that DNNs trained with the negative Pearson correlation coefficient as the loss function are particularly fit for photon-starved phase retrieval problems, which are fundamentally affected by strong Poison noise. In this paper we demonstrate that the use of a perceptual loss function significantly improves the reconstructions.
\end{abstract}

\section{Introduction}

In the last few years, the importance of deep learning in the field of computational imaging has been steadily growing. Deep neural networks (DNNs)\cite{nn:lecun15-dl} are used for a variety of tasks such as denoising~\cite{inv:remez17}, super-resolution imaging~\cite{inv:ledig17, inv:dong15-super-res, inv:dong14-super-res, inv:perceptual-loss,deng2018learning}, design of optimal illumination~\cite{horstmayer:2017}, optical and X-ray tomography~\cite{kamilov:2015, kamilov:2016, jin:2017, gupta:2018, inv:nguyen18-tomo}. The success of DNNs comes from their versatility and execution speed once they have been trained. They have proven particularly adapted for underdetermined and ill-posed problems, which were often solved using optimization methods that included some regularization scheme. In classical methods (other than DNNs), a regularizer is included in the function to optimize, its role being to include prior knowledge about the class of objects in the particular problem considered. The regularizer has to be designed to favor solutions that match our expectations about the object, for example smoothness, edge preservation, positivity, real-valuedness, geometrical support etc.. The task of designing the optimal regularizer is difficult for two reasons, first one may not know exactly what object features really matter for the problem considered, secondly even if these features were known, designing the proper regularization operator that favors them is not trivial. DNNs offer the significant advantage over regularized optimization that they \textit{learn} the prior in the data.

A classical computational imaging problem to which DNNs have been successfully applied is phase retrieval~\cite{metzler:2018, kemp:2018, boominathan:2018}, which can be stated as retrieving a possibly complex function from the modulus of its Fourier transform, and variations over that theme. As any other imaging problem, phase retrieval becomes increasingly difficult to solve as the photon budget available for the measurement gets reduced. In a recent paper~\cite{inv:goy2018low}, we demonstrated a method for phase retrieval in extremely low light conditions that combines a physics-inspired preprocessing step with a DNN. The preprocessing consists in performing an approximate projection the measurement back to the object plane so as to obtain a first guess of the object, hereafter called the `approximant.' Note that an exact projection is not possible as the phase is unknown. The role of the subsequent DNN is to both denoise the approximant and correct the distortion left over by the approximate projection. In this first work, the DNN was trained by maximizing the Pearson correlation coefficient (PCC) between the ground truth and the DNN output, the choice of this metric being motivated by empirical results obtained earlier~\cite{li:2018}.

In this paper, we show improved results over prior work for the photon-starved phase retrieval problem by optimizing the DNN with a perceptual loss function~\cite{inv:perceptual-loss}. The underlying idea of perceptual loss is to assess the quality of DNN reconstructions not by using an arbitrary metric, such as PCC or mean square error (MSE), but rather by the ability of the reconstruction to be successfully recognized in a classification task. In other words, we train the DNN to produce images that match the expectation of a trained classification algorithm about what valid images should look like. As we demonstrate here, the consequence of this design is that, for a human observer, the reconstructions display a much improved quality over PCC or MSE-trained DNN reconstructions. 

\section{Methods}

\subsection{Problem formulation}
In general, a noiseless computational imaging problem may be formulated as follows:
\begin{equation}
	g = H(f)
\end{equation}
\noindent where $f$ is the function (possibly complex) describing the object, $g$ the raw measurement (or collection of measurements) and $H$ a possibly nonlinear forward operator mapping $f$ to $g$. In the phase retrieval problem we consider here, $H$ is given explicitly by:
\begin{equation}
	H_{\mathrm{noisy}}(f) = \mathrm{Poisson}\left[\left|F\left(u_{\mathrm{inc}}e^{jf}\right)\right|^2\right] + \eta,
\end{equation}
\noindent where $F$ is the Fresnel operator describing the field propagation from the object plane to the detector plane, $u_{\mathrm{inc}}$ the complex field incident on the object, `Poisson' represents the Poisson process including the proper normalization constant for the incident light intensity, and $\eta$ is an additive Gaussian noise representing detection noise. In what follows, it is more convenient to define $H$ as the noiseless operator:
\begin{equation}
	H(f) = \left|F\left(u_{\mathrm{inc}}e^{jf}\right)\right|^2.
\end{equation}
Before we apply any DNN, we preprocess the measurement in order to form an approximant. This type of preprocessing has been shown~\cite{inv:goy2018low, gupta:2018, inv:Sun2018, inv:nguyen18-tomo} to improve the DNN performance over an end-to-end DNN that would map the measurement directly to the object. We use a single step of the well-known Gerchberg-Saxton algorithm~\cite{gerchberg:1972} to perform this initial inversion~\cite{inv:goy2018low}:
\begin{equation}
	\tilde{f} = \arg\left\{F^{-1}\left(\sqrt{g}\arg\left\{F\left(u_{\mathrm{inc}}\right)\right\}\right)\right\},
\end{equation}
\noindent where $\tilde{f}$ represents the approximant and `arg' the argument or phase of the complex field. In a second step, a DNN is trained to map the approximant to an estimate $\hat{f}$ of the object:
\begin{equation}
	\hat{f} = \mathrm{DNN}\left(\tilde{f}\right).
\end{equation}
\subsection{VGG16 \cite{simonyan2014very} based Perceptual Loss as Training Loss}
It was long realized that pixel-wise losses (\textit{e.g.} MSE, MAE) generally encourage finding pixel-wise averages of plausible solutions which tend to be oversmooth \cite{gupta2011modified,wang2004image,wang2003multiscale}. Instead, perceptual loss, based on high-level image feature representations extracted from pre-trained CNNs, can be used as the loss function to generate images with good visual quality. It was first applied to various applications in image processing, including feature inversion\ \cite{mahendran2015understanding}, feature visualization \cite{simonyan2013deep,yosinski2015understanding}, and texture synthesis and style transfer\cite{gatys2015texture}. Later, \cite{johnson2016perceptual} first combined the advantages of the feed-forward neural networks and perceptual loss for style-transfer and super-resolution. To our knowledge, perceptual loss has not been successfully applied to phase retrieval. 

More specifically, the VGG16 network\cite{simonyan2014very} is a deep convolutional neural network (CNN) that has been proved successful on classification tasks on ImageNet\cite{imagenet:2009}. As the input images proceed through the VGG network layers, features relevant for the classification get progressively identified. By extracting the data at some depth within the network, we can obtain a collection of feature maps, which are expected to be a sparse representation of the object in the space of features relevant for classification. In \cite{johnson2016perceptual}, authors demonstrated generally high-level feature maps of VGG16 network contain content-related knowledge of the input examples, while low-level feature maps contain style information of the examples. Therefore, given an inverse problem on ImageNet, if the DNN is trained not by the conventional pixel-wise loss between the reconstructions and their corresponding ground truth examples, but instead by the loss of their corresponding feature maps at a certain high-level layer, then it would encourage the reconstructions generated to be ones that are more likely to be correctly classified by the VGG16. If we assume that human visual perception is associated with recognition and classification tasks, then we can expect that a DNN trained with the perceptual loss function will produce images that are of a better visual quality to a human observer. Therefore, VGG-based perceptual loss has been used extremely widely to enhance the visual quality of reconstructions. From a different point of view, the semantic knowledge learned from pre-trained examples tends to provide priors about the examples to be reconstructed which helps compensate the ill-posedness of the forward operator. Empirically, the loss function is formed on the feature maps extracted at a particular layer of VGG16 network, \textit{i.e.} \textbf{relu 2-2}, as is common practice in many prior works\cite{johnson2016perceptual,inv:ledig17}. 

Mathematically, suppose $n_{\mathrm{feat}}$ be the number of feature maps and $N_x\times N_y$ be the size of each feature map at layer \textbf{relu 2-2} of VGG16 network\cite{simonyan2014very}, then the VGG16 based perceptual loss between the ground truth $f$ and the reconstruction $\hat{f}$, is
\begin{equation}
\mathscr{L}(f,\hat{f})=\frac{1}{n_{\mathrm{feat}} N_x N_y}\sum_{i=1}^{n_{\mathrm{feat}}}\left|\left|\text{VGG}_{i}(f)-\text{VGG}_{i}(\hat{f})\right|\right|_{2}^{2},
\label{VGG-loss}
\end{equation}
where $\text{VGG}_{i}(l)$ denotes the $i^{\text{th}}$ feature map generated when passing image $l$ up to layer \textbf{relu 2-2} of VGG16. To cope with the dynamic range of the pre-trained VGG16 network, images need to be normalized to [-1,1] before entering into the pre-trained VGG16 network. 

\subsection{DNN design and training}

\begin{figure}
    \centering
    \includegraphics{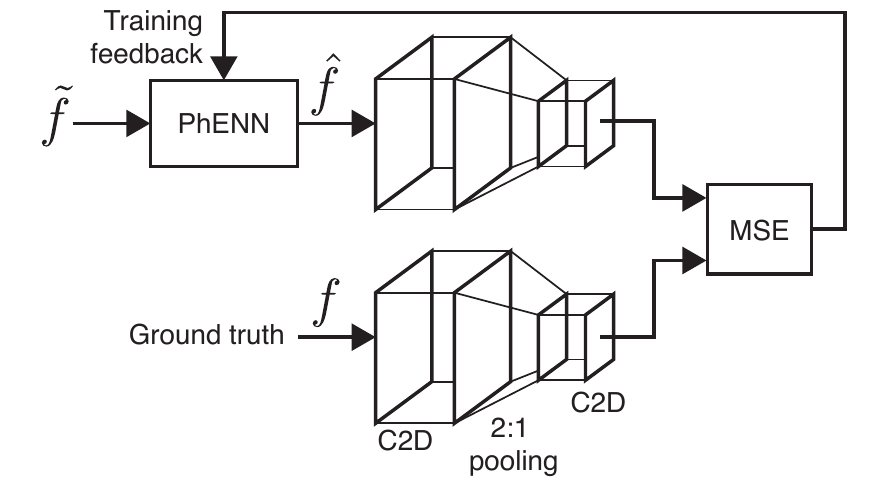}
    \caption{Diagram showing the principle of training the deep neural network for phase extraction (PhENN)\ \cite{inv:sinha17-PhENN} with perceptual loss function. The PhENN takes the approximant\ \cite{inv:goy2018low} $\tilde{f}$ as input. The estimate from PhENN $\hat{f}$ and the ground truth $f$ are both passed through the first following layers of the VGG16 DNN: two 2D convolution layers (denoted by C2D), a two-fold reduction in the lateral size performed by maximum pooling over 2 by 2 pixels groups, followed finally by another two 2D convolution layers. The two VGG16 outputs are compared by mean square error (MSE), which is used as the loss for the training of PhENN\ \cite{simonyan2014very}.}
    \label{fig:DNN}
\end{figure}

In what follows, we use a particular type of DNN that we developed earlier for phase extraction problems\ \cite{inv:sinha17-PhENN}. We denote it by PhENN for phase extraction neural network. The PhENN takes in the phase approximant $\tilde{f}$\ \cite{inv:goy2018low} as input and generates the phase estimate $\hat{f}$ , which is passed into the pre-trained VGG16 network up to layer \textbf{relu 2-2}, where the feature maps are compared with those generated by VGG16 from the ground truth examples $f$ at the same layer (Fig.\ref{fig:DNN}). 

The training examples are extracted from the benchmark ImageNet\ \cite{imagenet:2009} database. A set of 10,000 examples is extracted and split into three sets: a training set containing 9,500 examples, a validation set containing 450 examples, and a test set with $N_{\mathrm{test}} = 50$ examples, which will be used for the purpose of displaying the results and discussing performances. The DNN is trained on perceptual loss as mentioned in section 2.2 for 20 epochs. The computation is conducted on Tensorflow. 

\subsection{Calibration of reconstructions}\label{sec:cal}

The reconstruction by the perceptual loss trained neural networks is generally a nonlinear function of the corresponding ideal phase. For phase retrieval, we aim at producing a quantitatively accurate estimate of the phase. To that end, we perform a polynomial fit between the raw reconstructions and the ground truths from the validation set. We empirically tested polynomials with degrees ranging from 1 to 10 and found that polynomials with degree 3 were the best for this fit. We present the results from this calibration step later in Section 3.  


\section{Results}
\begin{figure}[ht]
    \centering
    \includegraphics[width=\textwidth]{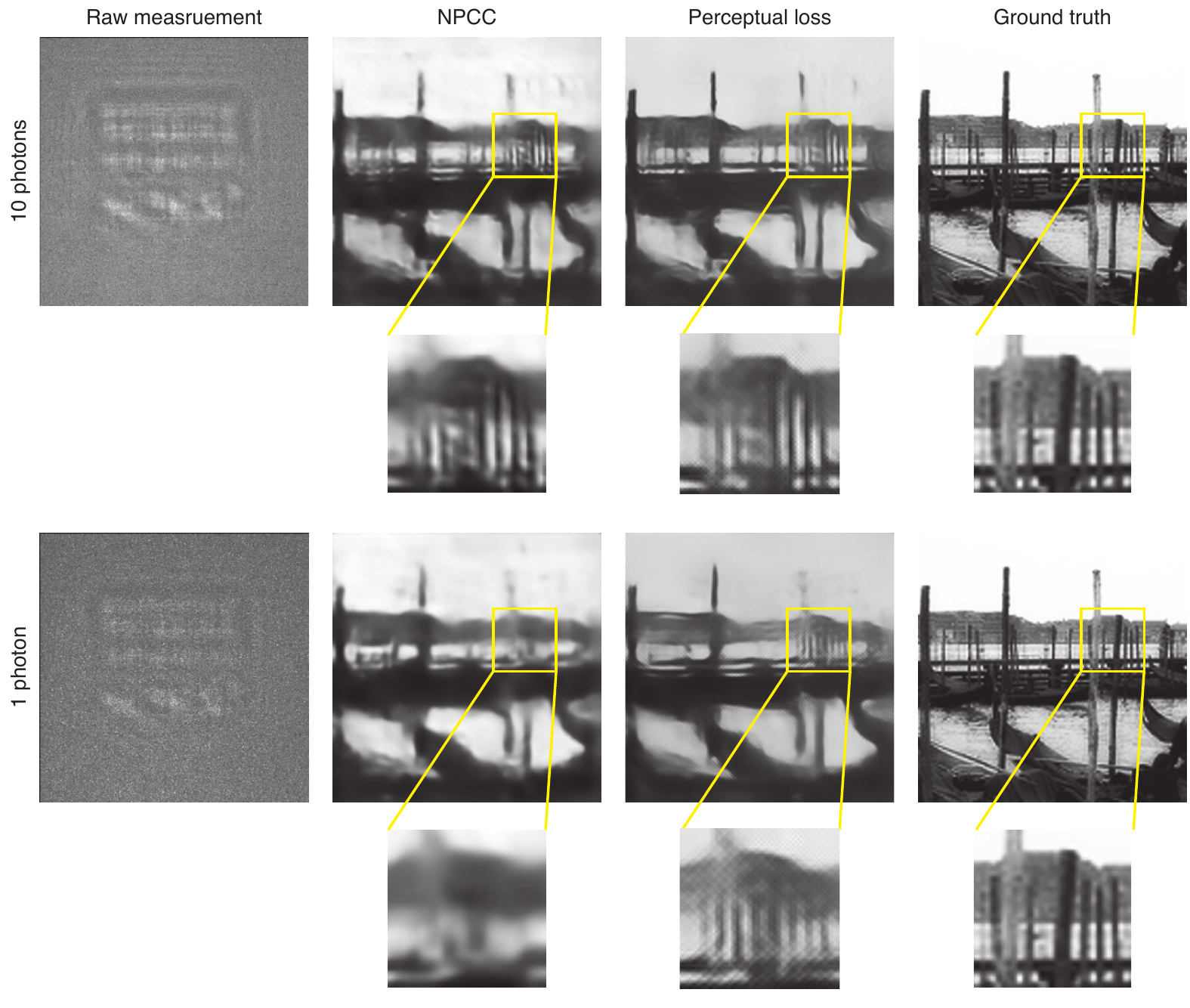}
    \caption{Comparison of reconstructions from PhENN trained with perceptual loss \textit{vs.} NPCC for 1 and 10-photon levels. The scaled up images show that some details are not rendered by the NPCC-trained PhENN whereas they become clearly identifiable with the perceptual loss function.}
    \label{result-boat}
\end{figure}

\begin{figure}[ht]
    \centering
    \includegraphics[width=\textwidth]{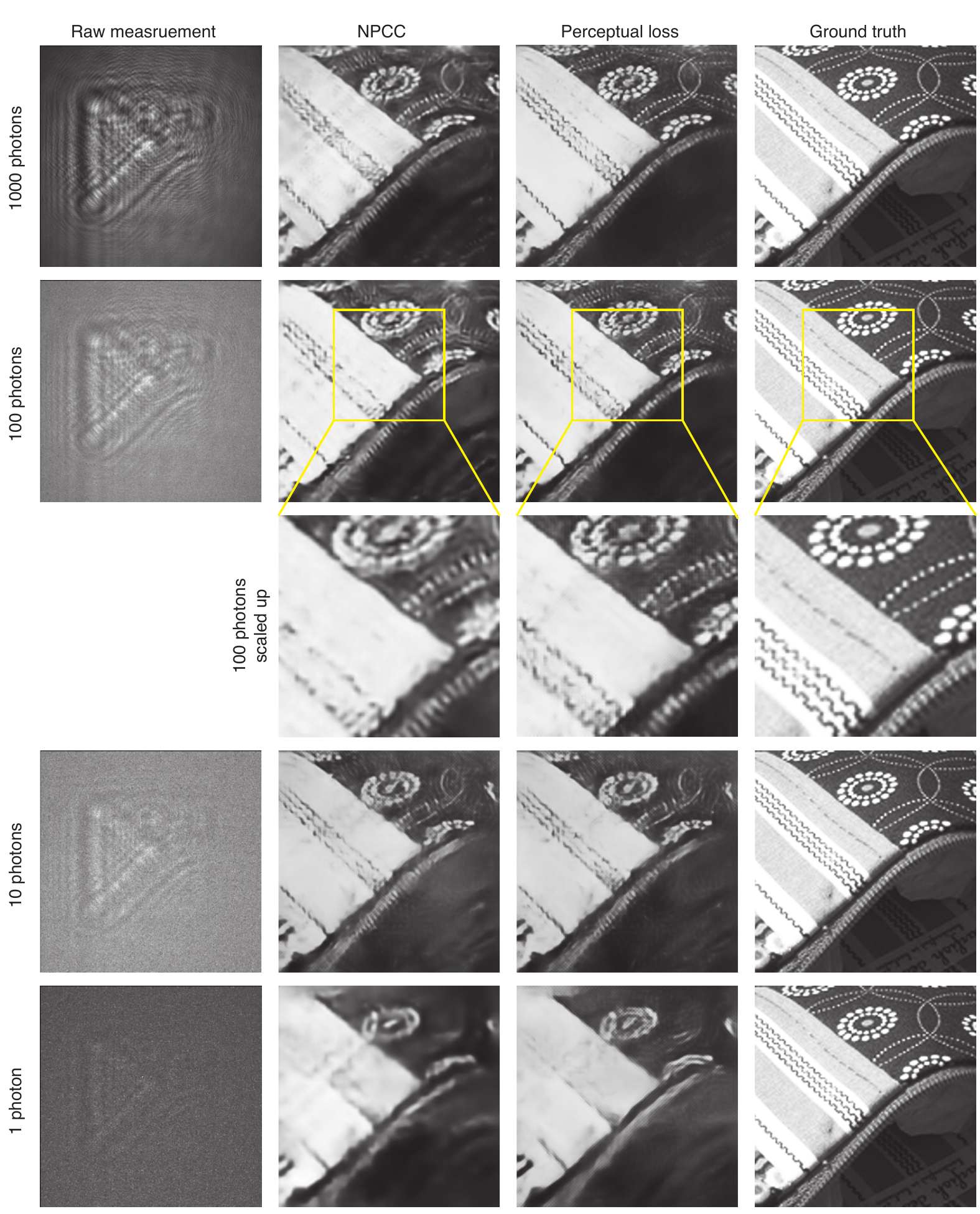}
    \caption{comparison of reconstructions from PhENN trained with perceptual loss \textit{vs.} NPCC for 1, 10, 100 and 1000-photon levels. In some areas, as shown by the scaled up images, some details are only visible in the perceptual loss reconstruction.}
    \label{result-art}
\end{figure}

In Fig.~\ref{result-boat}, we show comparisons of phase retrieval results for 1-photon and 10-photon levels, respectively, when the DNN is trained with perceptual loss, against those trained with Negative Pearson Correlation Coefficients (NPCC), as in \cite{inv:goy2018low}. In the case of perceptual loss, the reconstructions display sharper details in general and, as can be seen in the scaled up images in Fig.~\ref{result-boat}, show particular details such as the vertical posts in the scene, that are completely blurred in the NPCC image at 1-photon level. In Fig.~\ref{result-art}, we show a similar comparison for 4 different photon levels. The perceptual loss reconstructions look generally sharper and they show clearly more recognizable details, such as the wavy decoration patterns on the fabric. From the reconstructions, we see that the fine details (features) that would semantically help the classification get better recovered by the perceptual-loss trained DNN.

\section{Discussions on the artifacts in reconstructions}
\begin{figure}[ht]
    \centering
    \includegraphics[width=\textwidth]{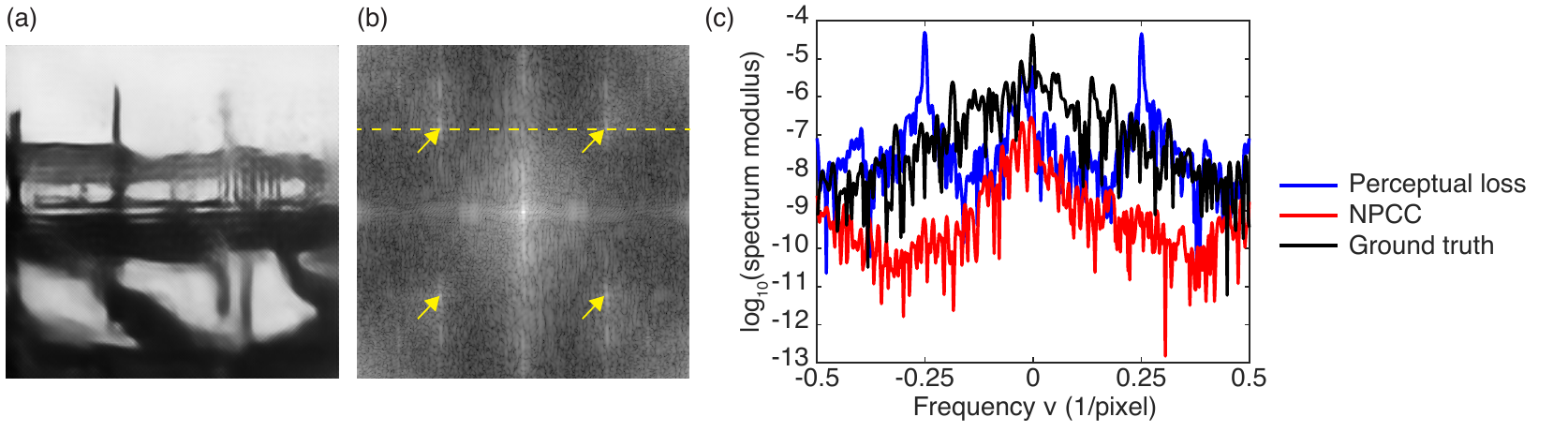}
    \caption{(a) Reconstruction by the perceptual loss trained DNN for the 1-photon level. (b) Log-scale image of the power spectral density of the reconstruction shown in (a). The artifact contributes in the modes indicated by the arrows. (c) Comparison between cross sections of the perceptual loss reconstruction power spectral density in blue, corresponding to image (b), with the power spectral density of the ground truth (black) and the power spectral density of the NPCC-trained DNN reconstruction (red).}
    \label{fig:2d-spec}
\end{figure}

The perceptual loss training provides significant improvement over the benchmark (the reconstructions obtained from NPCC-trained DNNs). However, at low photon levels, grid-like artifacts start appearing in the reconstructions, both before and after the calibration step described in section~\ref{sec:cal}. The artifacts are always centered at the same spatial frequency corresponding to a spatial period of 4 pixels, which we will refer to as the `characteristic frequency' $\nu_c$. In Fig.~\ref{fig:2d-spec}a, we show an example severely affected by the artifact and in Fig.~\ref{fig:2d-spec}b, which shows its spectral power density, the artifact is clearly visible at frequencies $(\nu_x, \nu_y) = (\pm \nu_c, \pm \nu_c)$. In Fig.~\ref{fig:2d-spec}c, we compare the cross sections of the spectra of the ground truth, the NPCC reconstruction, and the perceptual loss reconstruction respectively. The perceptual loss reconstruction spectrum displays a sharp peak at frequency $\nu_c$ that is not present in the ground truth, or the NPCC reconstructions.

\begin{figure}[ht]
    \centering
    \includegraphics[width=\textwidth]{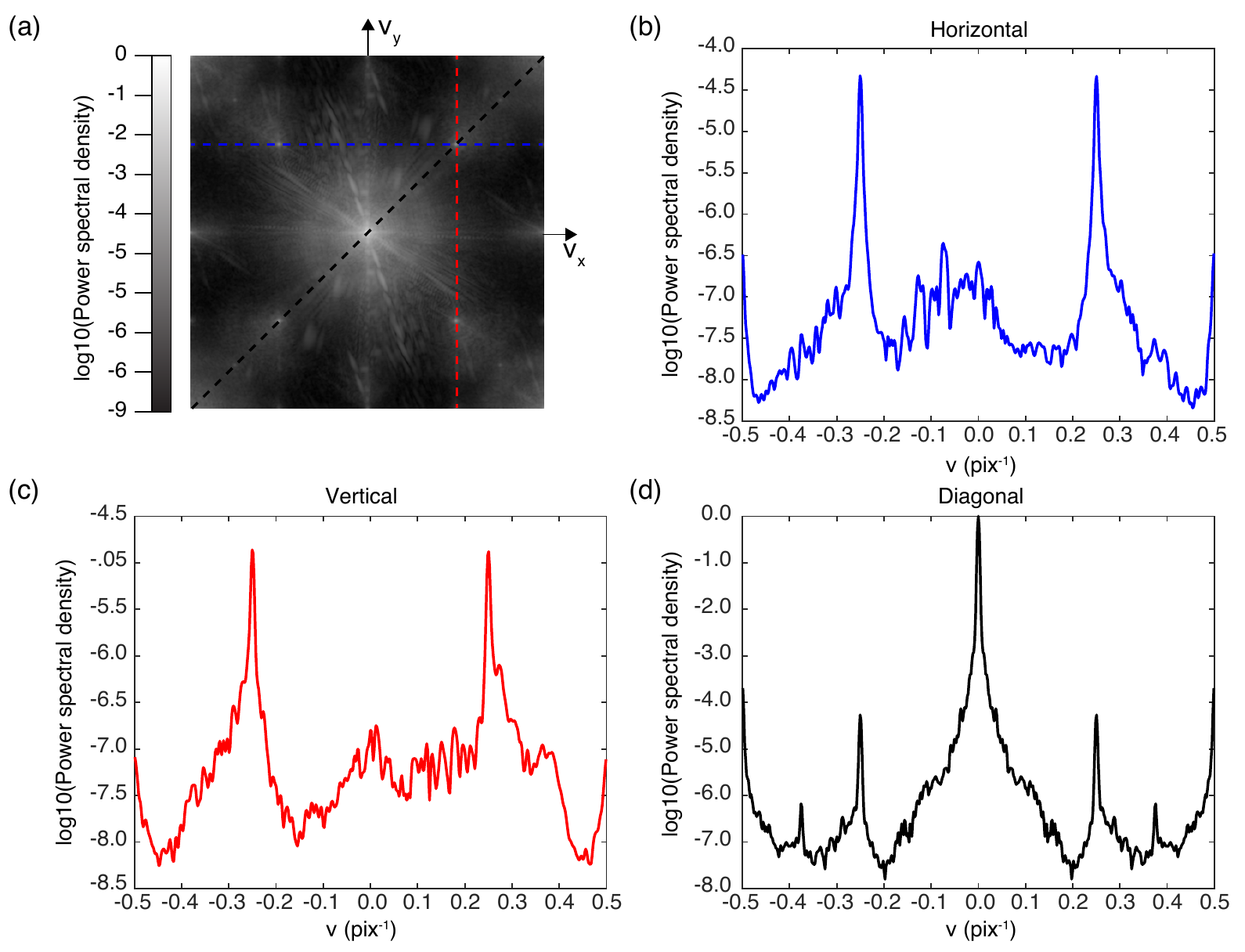}
    \caption{(a) Log-amplitude of the power spectral density of each reconstruction $\hat{f}$ clearly showing the signature of the artifact, which shows as a prominent network of horizontal and vertical strips. (b) Horizontal profile of (a). (c) Vertical profile of (a). (d) Diagonal profile of (a).}
    \label{fig:rec_psd}
\end{figure}

The characteristic signature of the artifact is even more pronounced if we consider the average of all 50 examples in the test set. In Fig.~\ref{fig:rec_psd}(a), we show the amplitude of the average spectrum (2D power spectral density) of all test set reconstructions. In what follows, we investigate VGG's particular behavior at the characteristic frequency and provide some insights into the cause of grid-like artifacts.

The fact that the artifacts only occur in cases where the noise is strong, explains why it has not been reported before to our knowledge. This motivates us to investigate more deeply on the origin of this artifact and its connection to the VGG network.

\subsection{VGG16's effect on the characteristic frequency}\label{sec:vgg_effect1}

To investigate the formation of such artifacts and more specifically whether the pre-trained VGG treats the characteristic frequency any differently from others, we conduct the following test: for each ground truth image $f_{i}$ in the test set, we generate a noisy version of it by adding noise at a particular spatial frequency $(\nu_{x0}, \nu_{y0})$ with amplitude $A$, which is empirically pre-defined to be 0.1. Thus, the strength of the artifacts in the noisy images are visually comparable with those in the reconstructions at 1 photon. More specifically, we define the noise as:
\begin{equation}
\begin{aligned}
    \xi(A,x, y, \nu_{x0}, \nu_{y0}) =  A \mathscr{F}^{-1}\{ & e^{ja}\delta(\nu_x-\nu_{x0}, \nu_y-\nu_{y0}) + e^{-ja}\delta(\nu_x+\nu_{x0}, \nu_y+\nu_{y0}) \\ + & e^{jb}\delta(\nu_x+\nu_{x0}, \nu_y-\nu_{y0}) +   e^{-jb}\delta(\nu_x-\nu_{x0}, \nu_y+\nu_{y0}) \},
\end{aligned}
\label{syn-noise}
\end{equation}
\noindent where $\mathscr{F}$ is the Fourier transform, $\delta$ the Dirac impulse, and $a$ and $b$ two random real numbers uniformly distributed in $[-\pi, \pi]$. The noisy and clean images satisfy: 
\begin{equation}\label{eqn:noise}
f_{\text{noisy},i}(A, x, y, \nu_{x0},\nu_{y0})= f_{i} + \xi(A, x, y,\nu_{x0},\nu_{y0})
\end{equation}
We then submit the clean set $F=\{f_i, i=1,\cdots, N_{\mathrm{test}}\}$ and the corresponding noisy set, $F_{\text{noisy}}(A,\nu_{x0},\nu_{y0})=\{f_{\text{noisy}, i}(A,\nu_{x0},\nu_{y0}), i=1,\cdots, N_{\mathrm{test}}\}$, into the pre-trained VGG16 up to layer \textbf{relu2-2} and compute the sum of the losses for all examples $i$:
\begin{equation}\label{eq:perc_loss}
\mathscr{L}(f,f_{\text{noisy}})=\frac{1}{N_{\text{test}}}\sum_{i=1}^{N_{\mathrm{test}}}\mathscr{L}(f_i,f_{\text{noisy},i})
\end{equation}
The loss, $\mathscr{L}(f,f_{\text{noisy}})$ characterizes how the pre-trained VGG16 reacts to disparity at frequency $(\nu_{x0},\nu_{y0})$. We scan the whole Fourier plane and compute a loss according to each frequency to understand responses of pre-trained VGG16 to disparities at all possible frequencies. 

\begin{figure}[ht]
    \centering
    \includegraphics[width=\textwidth]{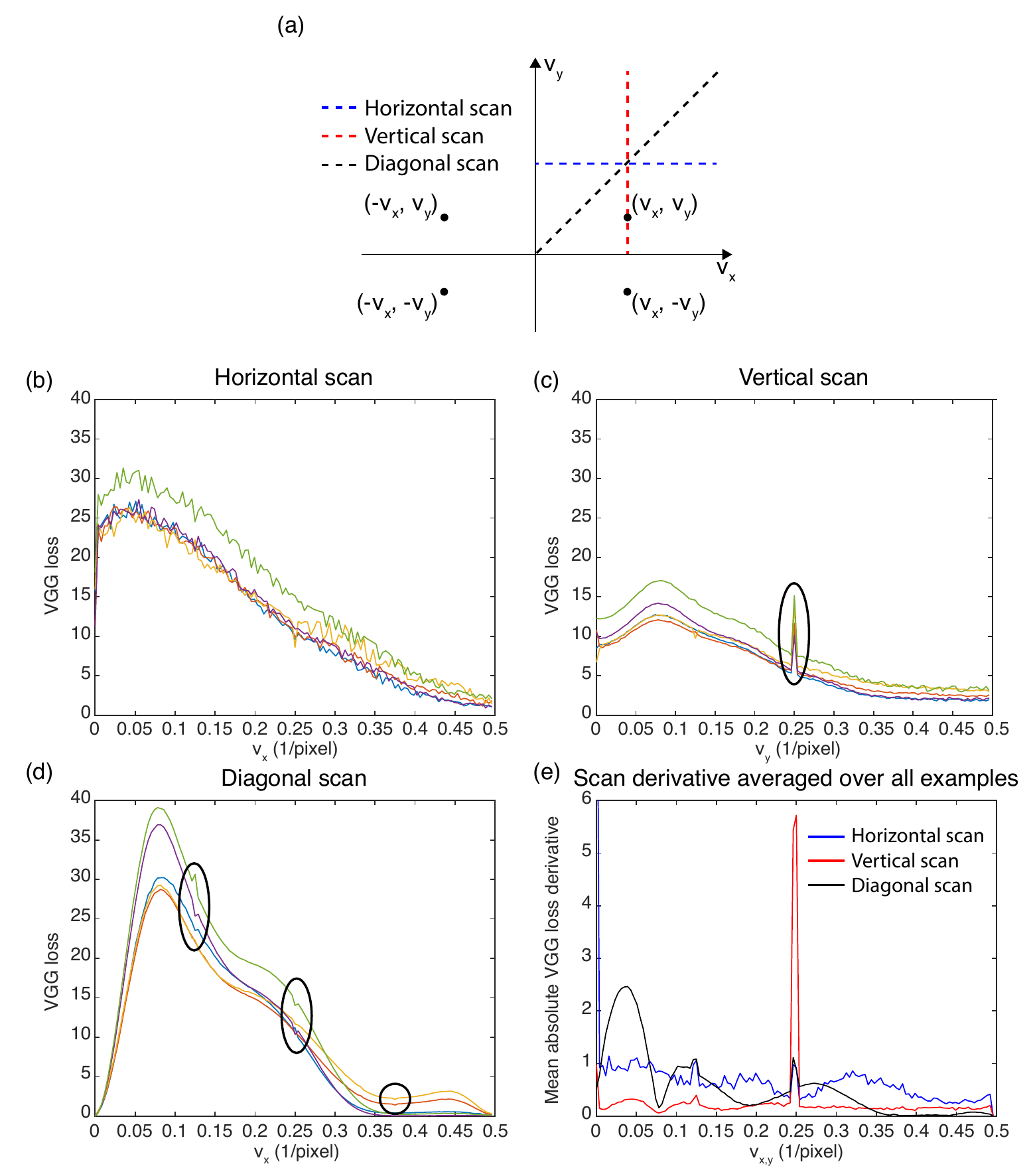}
    \caption{Dependence of VGG16 loss on the frequency of the noise. (a) Diagram showing the scanning scheme in the Fourier domain.  The noise $n$ is added on at a single frequency and made Hermitian, \textit{i.e.} $n(\nu_x, \nu_y) = n(-\nu_x, -\nu_y)^*$. (b) Loss as a function of frequency for the horizontal scan and five examples from the test set, for a noise amplitude of $A = 0.1$. (c) Loss as a function of frequency for the vertical scan for the same five examples. (d) Loss as a function of frequency for the diagonal scan for the same five examples. (e) Absolute value of the derivative of the loss with respect to frequency. The values are averaged over the 50 examples of the test set and plotted for the horizontal, vertical and diagonal scans. The ellipses in (c) and (d) indicate where strong non-smoothness can be observed in the loss curves. The position of the spikes correspond to artifact features observed in the spectrum of the average reconstruction.}
    \label{fig:2d-spec-scan}
\end{figure}

Here, we only show frequency scans along three representative directions, \textit{i.e.} horizontal, vertical and diagonal (Fig.~\ref{fig:2d-spec-scan}a). In Fig.~\ref{fig:2d-spec-scan}b-e, we show the corresponding profiles, for five randomly picked examples. We see that in all three scanning directions, for a majority of examples, there is a non-smoothness at the characteristic frequency $0.25\text{pixel}^{-1}$. The horizontal and vertical profiles display significantly different shapes (and magnitudes of the non-smoothness), which indicates that the convolution filters in the pretrained VGG are not symmetric. This can be expected from the fact that, in the classification task for which VGG was trained, discrimination of image orientation is important. In Fig.~\ref{fig:2d-spec-scan}, we circled the portion of the loss curves where strong non-smoothness occurs. The magnitude and sign (\textit{i.e.} whether it is a positive of negative fluctuation) of the non-smoothness vary from example to example, therefore, we consider the mean absolute derivative of the loss function as a suitable metric to detect the non-smoothness (Fig.~\ref{fig:2d-spec-scan}e).

While we do not expect a perfect match between the reconstruction spectrum of Fig.~\ref{fig:rec_psd}b-d and the VGG frequency response in Fig.~\ref{fig:2d-spec-scan}b-d, we still expect that VGG displays a particular behavior at the characteristic frequency and that is, indeed, what we observe. Knowing that, in what follows, we investigate more deeply the mechanism of why the  perceptual loss based training leads to the survival of the artifact at the characteristic frequency.


\subsection{Minimization of the perceptual loss} \label{sec:minimize AP}

The results in the previous section suggest that the VGG network is primarily responsible for the appearance of the artifact. A common way to investigate the internal mechanism of a neural network is to compute so-called maximally activated patterns (MAPs) \cite{zeiler2014visualizing}. The idea is to find, through optimization, the input to the network that would maximize some metric defined on a given layer within the network. MAPs are thus functions of the particular layer on which they are defined. For the layer of interest, the MAP represents what the layer is most sensitive to. In a classification network, such as VGG, MAPs suggest what patterns may contribute most to the success of the classification. In the perceptual loss training, we consider layer \textbf{relu2-2}, we are thus interested in the MAP defined for that particular layer.

Formally, we use the following definintion of the MAP, based on the norm of the feature maps at layer \textbf{relu2-2}:

\begin{equation}\label{eq:map}
\text{MAP}=\underset{\eta}{\operatorname{argmax}} \left\{\|\text{VGG}(\eta)\|\right\} \text{\ such\ that\ } \eta_p\in[0, 1]
\end{equation}
\noindent where VGG stands for the mapping from an image to the VGG \textbf{relu 2-2} layer, and $\eta_p$ the pixels of $\eta$.

MAPs provide a methodology to study the response of DNNs to their inputs, and can suggest what input patterns may get amplified or suppressed through the network. Because of the possibly strong nonlinearity of the network, we suggest to consider the response of the \textbf{relu2-2} not with respect to the whole input itself (which would be the MAP defined in Eq. \ref{eq:map}), but rather with respect to perturbations added on top of input images from the ImageNet dataset. We propose to find out what perturbations are left over after a minimization of the perceptual loss $\mathscr{L}$ defined in Eq. \ref{eq:perc_loss}. We expect that the artifact typically observed in the perceptual loss trained PhENN reconstructions lie in the set of perturbations only weakly affected by the minimization of the perceptual loss.

To that end, we perform numerical tests by initializing the minimization algorithm with a noisy version $f_{\text{noisy}}(\nu_{x0}, \nu_{y0})$ of the ground truth on which noise at a particular spatial frequency (as defined in Eq.~\ref{eqn:noise}) has been added. That is:
\begin{equation}\label{eq:min_vgg_map}
\hat{f}=\underset{\eta}{\operatorname{argmin}}\left\{\mathscr{L}\left(\eta, f\right)\right\}.
\end{equation}
This minimization problem, because it is initialized with $f_{\text{noisy}}(\nu_{x0}, \nu_{y0})$, implicitly defines an operator from the noise $\xi(\nu_{x0}, \nu_{y0})$ to $\hat{f}$, which we can write as:
\begin{equation}\label{eq:min_vgg_map2}
\hat{f}(\nu_{x0}, \nu_{y0})=G_f[\xi(\nu_{x0}, \nu_{y0})].
\end{equation}

\begin{figure}[ht]
    \centering
    \includegraphics{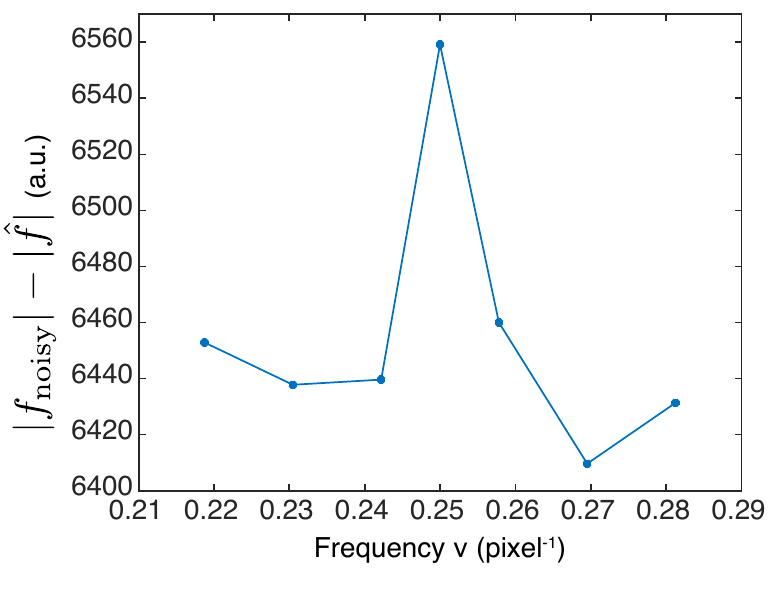}
    \caption{Change in the frequency components of an image through the minimization operation of Eq.~\ref{eq:min_vgg_map}. The frequency $\nu$ refers to position $(\nu_x, \nu_y)=(\nu, \nu)$ in the Fourier domain (diagonal scan). The value plotted is the difference of the modulus of the spectrum of the noisy image $\tilde{f}$ (defined in Eq.~\ref{eqn:noise}) and the spectrum of $\hat{f}$, the result of optimization~\ref{eq:min_vgg_map} starting from $\tilde{f}$.}
    \label{fig:freq_map}
\end{figure}

\noindent One may think that problem~\ref{eq:min_vgg_map} necessarily converges to the ground truth (\textit{i.e.} $\hat{f}(\nu_{x0}, \nu_{y0})=f$ for all $(\nu_{x0}, \nu_{y0})$); however, due to the non-linearity in VGG16, it is not expected to be convex and may converge instead to a local minimum that depends on $(\nu_{x0}, \nu_{y0})$. We are interested in the following: at what frequency $(\nu_{x0}, \nu_{y0})$ does the noise get most reduced by minimizing VGG loss? In Fig.~\ref{fig:freq_map}, we show of this test. Consistent with the observations in section~\ref{sec:vgg_effect1}, the noise at the characteristic frequency undergoes the strongest suppression, which indicates that the disparity at the characteristic frequency would give rise to higher VGG loss than its neighbors.

Therefore, when the PhENN is trained to minimize the VGG-based perceptual loss, the training implicitly tends to match the reconstructions and the ground truth examples at the characteristic frequency more than its neighbors. Other frequencies may not need to be perfectly matched to achieve a low VGG loss, thus the training stagnates at some local minimum. At such local minimum, the characteristic frequency stands out due to deficiencies of its neighboring frequencies, manifesting as the artifacts centered at the characteristic frequency. Further verification of the conjecture requires an autopsy of the weights of the pre-trained VGG, which is beyond the scope of this paper.

\section{Conclusion and Future Works}

We demonstrated that the PhENN yields superior results when trained with a VGG-based perceptual loss than with the NPCC, which we considered the benchmark so far for phase retrieval. The reconstructions are, however, systematically impaired by an artifact localized, in the Fourier domain, at specific frequencies and mostly at a characteristic frequency $\nu_c = 0.25\text{pixel}^{-1}$. The fact that the artifact follows such a structured pattern suggests that it is induced by the VGG network and that it is not simply an example dependent side effect. The numerical experiments we conducted on VGG show that its frequency response share commonalities with the spectrum of the artifact, notably by the fact that non-smoothness is observed in the VGG spectrum at $\nu_c, \frac{1}{2}\nu_c$ and $\frac{3}{2}\nu_c$. Moreover, we showed that minimization of the perceptual loss \textit{per se} has uneven effect on the different frequencies of an image and that the typical artifact observed in the perceptual loss reconstruction survives the minimization process. The internal structure of VGG needs to be investigated further to identify the cause of this type of frequency dependent artifact.

In this paper, the question of what quantitative metric should be used to assess the quality of the reconstructions is of primary importance. It is commonly acknowledged that metrics such as Peak Signal-to-Noise Ratio (PSNR)~\cite{gupta2011modified,johnson2016perceptual,inv:ledig17}, Structural Similarity index (SSIM)~\cite{wang2003multiscale,wang2004image} do not reliably correlate with human visual perceptions, which is the reason why we refrained from drawing conclusions based on these metrics. If quantitative quality assessment is required, two avenues can be considered: 1- the quantification of the human assessment by a statistical study over an extensive population of human observers in normalized conditions. This quantification approach, which can be applied universally to any reconstruction method, gives the human perception the last word, a decision that obviously depends on the application. Another avenue is the use of the very definition of perceptual loss as based on a classification network, \textit{i.e.} the quality of a reconstruction would be associated to how well it can be classified by a well-established image classification neural network, such as the VGG. Indeed, the reconstruction is considered good when the features required for good classification have been properly reconstructed. We suggest that this is what is done implicitly by a human observer, \textit{i.e.} a human will grant an image a good score if its content can be recognized, that is classified, provided we consider that pattern recognition is the mapping of observed features to labelled objects. We consider that such studies are beyond the scope of this paper, which only aimed at demonstrating, visually, additional details could be retrieved by using the perceptual loss.

\section*{Acknowledgments}
This work was supported by the Intelligence Advanced Research Projects Activity (IARPA) grant No. FA8650-17-C-9113 and the SenseTime company.

\bibliography{2019Perceptualloss}

\end{document}